\documentclass[reprint,showpacs]{revtex4-1}
\pdfoutput=1
\usepackage[utf8]{inputenc}
\usepackage[T1]{fontenc}
\usepackage{mathtools}
\usepackage{bm}
\usepackage[unicode]{hyperref}
\usepackage{siunitx}
\usepackage{upgreek}
\usepackage{xspace}
\usepackage{placeins}

\usepackage[svgnames]{xcolor}
\usepackage[normalem]{ulem}

\newcommand{\upe}{\ensuremath{\mathrm{e}}}
\newcommand{\dif}{\ensuremath{\mathrm{d}}}
\newcommand{\kB}{\ensuremath{k_\mathrm{B}}}
\newcommand{\kBT}{\ensuremath{\kB T}}
\newcommand{\kBTe}{\ensuremath{\kB T_\mathrm{e}}}
\newcommand{\kBTm}{\ensuremath{\kB T_\mathrm{m}}}

\newcommand{\uv}[1]{\ensuremath{\mathbf{\hat{#1}}}}

\newcommand{\bose}[1]{\ensuremath{n_\mathrm{B}\left(#1\right)}}
\newcommand{\bosepow}[2]{\ensuremath{\left[n_\mathrm{B}\left(#2\right)\right]^#1}}

\hypersetup{
    colorlinks=false,
    pdfauthor={W.P. Sterk},
    pdfsubject={USMR},
    pdftitle={Magnon contribution to unidirectional spin Hall
    magnetoresistance},
    pdfkeywords={USMR,
    magnon,
    magnetoresistance}
}

\newcommand{\ITP}{Institute for Theoretical Physics, Utrecht University,
Princetonplein~5, 3584~CC Utrecht, The~Netherlands\xspace}

\begin{document}
\author{W.P. Sterk}
\affiliation{\ITP}
\author{D. Peerlings}
\affiliation{\ITP}
\author{R.A. Duine}
\affiliation{\ITP}
\affiliation{Department of Applied Physics, Eindhoven University of
Technology, PO~Box~513, 5600~MB Eindhoven, The Netherlands\xspace}
\title{Magnon contribution to unidirectional spin Hall
magnetoresistance in ferromagnetic-insulator/heavy-metal bilayers}
\date{\today}
\begin{abstract}
We develop a model for the magnonic contribution to the unidirectional spin
Hall magnetoresistance (USMR) of heavy metal/ferromagnetic insulator bilayer
films.  We show that diffusive transport of Holstein-Primakoff magnons leads
to an accumulation of spin near the bilayer interface, giving rise to a
magnoresistance which is not invariant under inversion of the current
direction. Unlike the electronic contribution described by Zhang and Vignale
[Phys. Rev. B \textbf{94}, 140411 (2016)], which requires an electrically
conductive ferromagnet, the magnonic contribution can occur in ferromagnetic
insulators such as yttrium iron garnet. We show that the magnonic USMR is, to
leading order, cubic in the spin Hall angle of the heavy metal, as opposed to
the linear relation found for the electronic contribution. We estimate that
the maximal magnonic USMR in Pt|YIG bilayers is on the order of $10^{-8}$, but
may reach values of up to $10^{-5}$ if the magnon gap is suppressed, and can
thus become comparable to the electronic contribution in e.g. Pt|Co. We show
that the magnonic USMR at a finite magnon gap may be enhanced by an order of
magnitude if the magnon diffusion length is decreased to a specific optimal
value that depends on various system parameters.
\end{abstract}

\maketitle

\section{Introduction}
The total magnetoresistance of metal/ferromagnet
heterostructures is known to comprise several independent contributions,
including but not limited to anisotropic magnetoresistance (AMR)
\cite{1975ITM....11.1018M}, giant magnetoresistance (GMR, in stacked magnetic
multilayers) \cite{1989PhRvB..39.4828B} and spin Hall magnetoresistance (SMR)
\cite{2013PhRvB..87n4411C}. A common characteristic of these effects is
that they are linear; in particular, this means the measured magnetoresistance
is invariant under reversal of the polarity of the current.

In 2015, however, \citet{2015NatPh..11..570A} measured a small but distinct
asymmetry in the magnetoresistance of Ta|Pt and Co|Pt bilayer films.  Due to
its striking similarity to the current-in-plane spin Hall effect (SHE) and
GMR, save for its nonlinear resistance/current characteristic, this effect was
dubbed unidirectional spin Hall magnetoresistance (USMR).

In the years following its discovery, USMR has been detected in bilayers
consisting of magnetic and nonmagnetic topological insulators
\cite{2016PhRvL.117l7202Y}, and the dependence of the USMR on layer thickness
has been investigated experimentally for Co|Pt bilayers
\cite{2017ApPhL.111w2405Y}. Additionally, \citet{2017ApPhL.110t3506A} have
shown that USMR may be used to distinguish between the four distinct magnetic
states of a ferromagnet|normal metal|ferromagnet trilayer stack, highlighting
its potential application in multibit electrically controlled memory cells.

Although USMR is ostensibly caused by spin accumulation at the
ferromagnet|metal interface, a complete theoretical understanding of this
effect is lacking. In bilayer films consisting of ferromagnetic metal (FM) and
heavy metal (HM) layers, electronic spin accumulation in the ferromagnet
caused by spin-dependent electron mobility provides a close match to the
observed results \cite{2016PhRvB..94n0411Z}. It remains unknown, however,
whether this is the full story; indeed, this model's underestimation of the
USMR by a factor of two lends plausibility to the idea that there may be
additional, as-yet unknown contributions providing the same experimental
signature. Additionally, the electronic spin accumulation model cannot be
applied to bilayers consisting of a ferromagnetic insulator (FI) and a HM, as
there will be no electric current in the ferromagnet to drive accumulation of
spin.

\citet{2016arXiv160308746K} have measured the USMR of Py|Pt (where Py denotes
for permalloy) bilayer and claim, using qualitative arguments, that a
\emph{magnonic} process is involved. Likewise, for Co|Pt and CoCr|Pt, more
recent results by \citet{2018arXiv180605305O} argue in favor of the presence
of a magnon-scattering contribution consisting of terms linear and cubic in
the applied current, and having a magnitude comparable to the electronic
contribution of \citet{2016PhRvB..94n0411Z}. Although these experimental
results provide a great deal of insight into the underlying processes, a
theoretical framework against which they can be tested is presently lacking.
In this work, we aim to take first steps to developing such a framework, by
considering an accumulation of magnonic spin near the FI|HM bilayer interface,
which we describe by means of a drift-diffusion model.

The remainder of this article is structured as follows: in
Sec.~\ref{sec:model}, we present our analytical model as generically as
possible. In Sec.~\ref{sec:results} we analyze the behavior of our model
using parameters corresponding to a Pt|YIG (YIG being yttrium iron garnet)
bilayer as a basis. In particular, in Sec.~\ref{subsec:eqt} we give
quantitative predictions of the magnonic USMR in terms of the applied current
and layer thicknesses, and in Sec.~\ref{subsec:thermal} we take into account
the effect of Joule heating. In the remainder of Sec.~\ref{sec:results}, we
investigate the influence of various material parameters. Finally, in
Sec.~\ref{sec:conclusion} we summarize our key results and present some open
questions.

\section{Magnonic spin accumulation}
\label{sec:model}
To develop a model of the magnonic contribution to the USMR, we focus on the
simplest FI|HM heterostructure: a homogeneous bilayer. We treat the transport
of magnonic and electronic spin as diffusive, and solve the resulting
diffusion equations subject to a quadratic boundary condition at the
interface. In this approach, valid in the opaque interface limit,
current-dependent spin accumulations---electronic in the HM and magnonic in
the FI---form near the interface. In particular, the use of a nonlinear
boundary condition breaks the invariance of the SMR under reversal of the
current direction, i.e. it produces USMR.

\begin{figure}
    \centering
    \includegraphics[width=\linewidth]{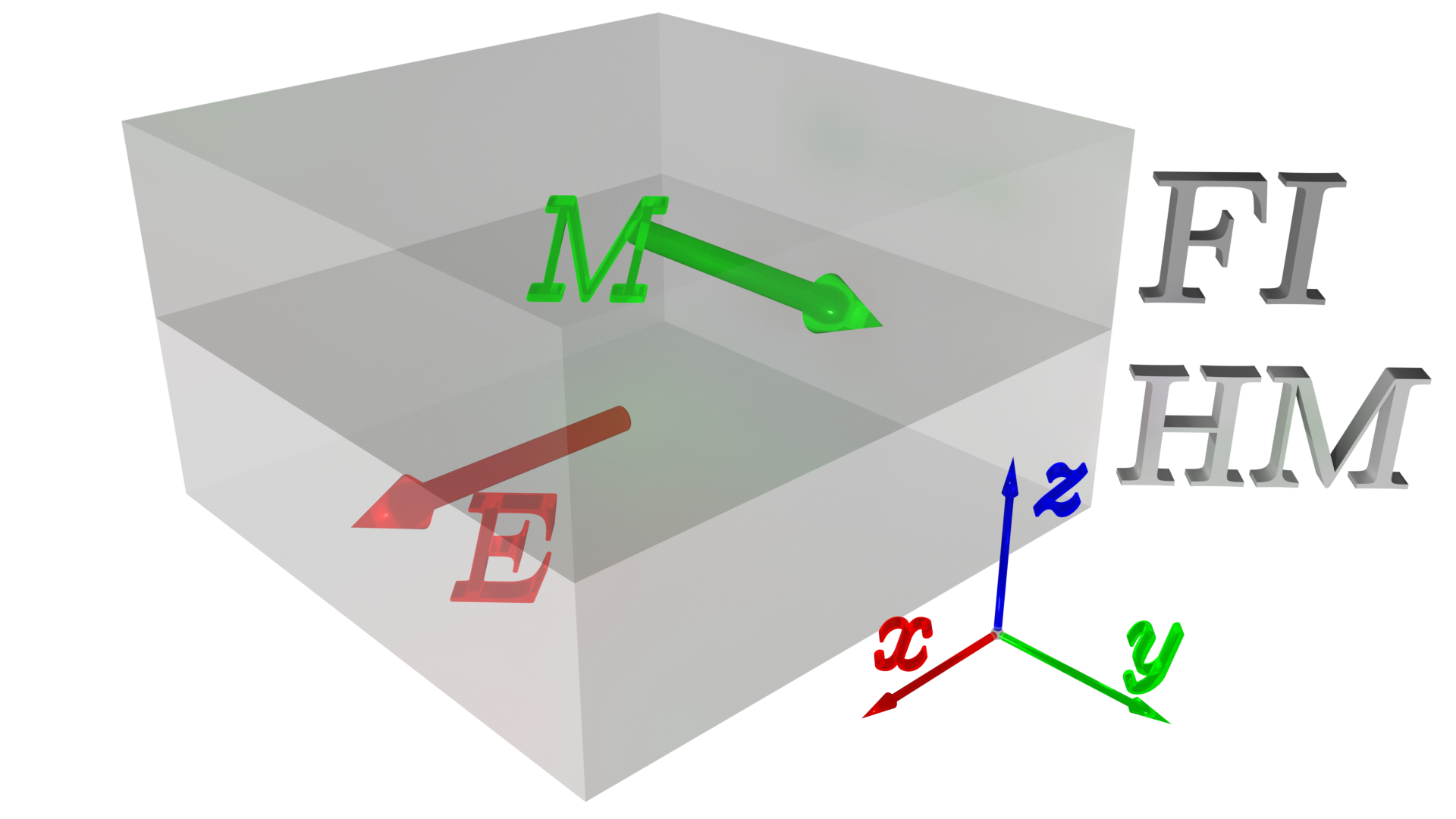}
    \caption{Schematic depiction of our system. The magnetization $M$ of the
    FI layer lies in the $+y$ direction, an electric field of
    magnitude $E$ is applied to the heavy metal layer (HM) in the $\pm x$
    direction, and the interface between the layers lies in the $xy$ plane.}
    \label{fig:system}
\end{figure}

We consider a sample consisting of a FI layer of thickness $L_\mathrm{FI}$
directly contacting a HM layer of thickness $L_\mathrm{HM}$.  We take the
interface to be the $xy$ plane, such that the FI layer extends from $z=0$ to
$L_\mathrm{FI}$ and the HM layer from $z=-L_\mathrm{HM}$ to 0. The
magnetisation is chosen to lie in the positive $y$-direction, and an electric
field $\bm{E}=\pm E\uv{x}$ is applied in the $x$-direction. The set-up is
shown in Fig.~\ref{fig:system}.

The extents of the system parallel to the interface are taken to be infinite,
and the individual layers completely homogeneous. This allows us to treat the
system as quasi-one-dimensional, in the sense that we will only consider spin
currents that flow in the $z$-direction. We account for magnetic anisotropy
only indirectly through the existence of a magnon gap. We further assume that
our system is adequately described by the Drude model (suitably extended to
include spin effects\cite{2007PhRvL..99t6601C}), and that the interface
between layers is not fully transparent to spin current, i.e., has a finite
spin-mixing conductance \cite{2015NatPh..11..496Z}. For simplicity, we assume
electronic spin and charge transport may be neglected in the ferromagnet, as
is the case for ferromagnetic insulators.

We describe the transfer of spin across the interface microscopically by the
continuum-limit interaction Hamiltonian\begin{align*}
    H_\mathrm{int}=-\int\dif^3\bm{r}\dif^3\bm{r'}\,J(\bm{r},\bm{r'})\Big[&
            b^\dagger(\bm{r'})c^\dagger_\downarrow(\bm{r})c_\uparrow(\bm{r})
            \nonumber \\
        &+b(\bm{r'})c^\dagger_\uparrow(\bm{r})
            c_\downarrow(\bm{r})\Big],
\end{align*}
where $c^\dagger_\alpha(\bm{r})$ [$c_\alpha(\bm{r})$] are fermionic creation
[annihilation] operators of electrons with spin
$\alpha\in\{\uparrow,\downarrow\}$ at position $\bm{r}$ in the HM, and
$b^\dagger(\bm{r'})$ [$b(\bm{r'})$] is the bosonic creation [annihilation]
operator of a circularly polarized Holstein-Primakoff magnon
\cite{1940PhRv...58.1098H} at position $\bm{r'}$ inside the ferromagnet. We
leave $J(\bm{r},\bm{r'})$ to be some unknown coupling between the electrons
and magnons, which is ultimately fixed by taking the classical limit
\cite{2015PhRvB..91n0402B,2002PhRvL..88k7601T}.

Transforming to momentum space and using Fermi's golden rule, we obtain the
interfacial spin current $j_\mathrm{s}^\text{int}$, which can be expressed in
terms of the real part of the spin mixing conductance per unit area
$g_\mathrm{r}^{\uparrow\downarrow}$ as
\cite{2012PhRvL.108x6601B,2015PhRvB..91n0402B}\begin{align}
    j_\mathrm{s}^\text{int}&=\frac{g^{\uparrow\downarrow}_\mathrm{r}}{\uppi s}
            \int\dif\varepsilon\,g(\varepsilon)(\varepsilon-\Delta\mu)
            \nonumber \\
        &\hspace{2em}\times\left[\bose{\frac{\varepsilon-\Delta\mu}{\kBTe}}
            -\bose{\frac{\varepsilon-\mu_\mathrm{m}}{\kBTm}}\right].
            \label{eq:intcur}
\end{align}
(Similar expressions were derived by \citet{2010JPhCS.200f2030T} and
\citet{2012PhRvB..86u4424Z}, although these are not given in terms of the
spin-mixing conductance.)

Here, $s$ is the saturated spin density in the FI layer, $g(\varepsilon)$ is
the magnon density of states, $\bose{x}=\left[\upe^x-1\right]^{-1}$ is the
Bose-Einstein distribution function, $\kB$ is Boltzmann's constant, and
$T_\mathrm{m}$ and $T_\mathrm{e}$ are the temperatures of the magnon and
electron distributions, respectively, which we do not assume \emph{a priori}
to be equal (although the equal-temperature special case will be our primary
interest). Of crucial importance in Eq.~(\ref{eq:intcur}) are the magnon
effective chemical potential $\mu_\mathrm{m}$---which we shall henceforth
primarily refer to as the magnon spin accumulation---and the electron spin
accumulation $\Delta\mu\equiv\mu^\uparrow-\mu^\downarrow$, which we define as
the difference in chemical potentials for the spin-up and spin-down electrons.
(In both cases, a positive accumulation means the majority of spin magnetic
moments point in the $+y$ direction.)

We employ the magnon density of states\begin{align*}
    g(\varepsilon)&=\frac{\sqrt{\varepsilon-\Delta}}{
            4\uppi^2J_\mathrm{s}^\frac{3}{2}}\Theta(\varepsilon-\Delta).
\end{align*}
Here, $J_\mathrm{s}$ is the spin wave stiffness constant, $\Theta(x)$ is the
Heaviside step function, and $\Delta$ is the magnon gap, caused by a
combination of external magnetic fields and internal anisotropy fields in
ferromagnetic materials \cite{2016ApPhL.108j2403T}. In our primary analysis of
a Pt|YIG bilayer, we take $\Delta\equiv\mu_\mathrm{B}\times\SI{1}{\tesla}
\approx\kB\times\SI{0.67}{\kelvin}$ with $\mu_\mathrm{B}$ the Bohr magneton,
in good agreement with e.g. \citet{1993PhR...229...81C}, and in
Sec.~\ref{subsec:gap} we specifically consider the limit of a vanishing
magnon gap.

To treat the accumulations on equal footing, we now redefine
$\mu_\mathrm{m}\to\delta\mu_\mathrm{m}$ and $\Delta\mu\to\delta\Delta\mu$,
expand Eq.~(\ref{eq:intcur}) to second order in $\delta$, and set $\delta=1$
to obtain\begin{align}
    j_\mathrm{s}^\text{int}\simeq-\Bigg[&\kBT_\mathrm{m}I_\mathrm{0}
            +I_\mathrm{e}\Delta\mu+I_\mathrm{m}\mu_\mathrm{m}
            +\frac{I_\mathrm{ee}}{\kBT_\mathrm{e}}(\Delta\mu)^2 \nonumber \\
        &+\frac{I_\mathrm{mm}}{\kBT_\mathrm{m}}\mu_\mathrm{m}^2
            +\frac{I_\mathrm{me}}{\kBT_\mathrm{m}}\mu_\mathrm{m}
            \Delta\mu\Bigg]\frac{g_\mathrm{r}^{\uparrow\downarrow}
            (\kBT_\mathrm{m})^\frac{3}{2}}{4\uppi^3 J_\mathrm{s}^\frac{3}{2}s}.
            \label{eq:intcur-exp}
\end{align}
Here, the $I_i$ are dimensionless integrals given by
Eqs.~(\ref{eq:intcur-coeffs}) in the Appendix. All $I_i$ are functions of
$T_\mathrm{m}$ and $\Delta$, and $I_0$, $I_\mathrm{e}$ and $I_\mathrm{ee}$
additionally depend on $T_\mathrm{e}$. In the special case where
$T_\mathrm{m}=T_\mathrm{e}$, $I_0$ vanishes, $I_\mathrm{m}=-I_\mathrm{e}$, and
$I_\mathrm{ee}=-(I_\mathrm{mm}+I_\mathrm{me})$.

In addition to $j_\mathrm{s}^\text{int}$, the spin accumulations and the
electric driving field $E$ give rise to the following spin currents in the $z$
direction:\begin{subequations}\label{eq:spincur}\begin{align}
    j_\mathrm{s}^\mathrm{e}&=\frac{\hbar}{2e}\left(-\frac{\sigma}{2e}
            \frac{\partial\Delta\mu}{\partial z}-\sigma\theta_\mathrm{SH}
            E\right), \label{eq:jse} \\
    j_\mathrm{s}^\mathrm{m}&=-\frac{\sigma_\mathrm{m}}{\hbar}
            \frac{\partial\mu_\mathrm{m}}{\partial z}. \label{eq:jsm}
\end{align}
\end{subequations}
Here $j_\mathrm{s}^\mathrm{e}$ and $j_\mathrm{s}^\mathrm{m}$ are the electron
and magnon spin currents, respectively. $\sigma$ is the electrical
conductivity in the HM, $\sigma_\mathrm{m}$ is the magnon conductivity in the
ferromagnet, $e$ is the elementary charge, and $\theta_\mathrm{SH}$ is the
spin Hall angle.

In line with \citet{2016PhRvB..94a4412C} and \citet{2012PhRvL.109i6603Z}, we
assume the spin accumulations $\mu_\mathrm{m}$ and $\Delta\mu$ obey diffusion
equations along the $z$-axis:\begin{align*}
    \frac{\dif^2\mu_\mathrm{m}}{\dif z^2}
        &=\frac{\mu_\mathrm{m}}{l_\mathrm{m}^2}, &
    \frac{\dif^2\Delta\mu}{\dif z^2}
        &=\frac{\Delta\mu}{l_\mathrm{e}^2},
\end{align*}
where $l_\mathrm{m}$ and $l_\mathrm{e}$ are the magnon and electron diffusion
lengths, respectively. We solve these equations analytically subject to
boundary conditions that demand continuity of the spin current across the
interface and confinement of the currents to the
sample:\begin{align*}
    j_\mathrm{s}^\mathrm{m}(0)&=j_\mathrm{s}^\mathrm{e}(0)
        =j_\mathrm{s}^\text{int}(0), \\
    j_\mathrm{s}^\mathrm{m}(L_\mathrm{FI})&=j_\mathrm{s}^\mathrm{e}(
            -L_\mathrm{HM})=0.
\end{align*}

This system of equations now fully specifies the magnonic and electronic spin
accumulations $\mu_\mathrm{m}$ and $\Delta\mu$, the latter of which enters the
charge current $j_\mathrm{c}$ via the spin Hall effect:\begin{align}
    j_\mathrm{c}(z)&=\sigma E+\frac{\sigma\theta_\mathrm{SH}}{2e}\frac{\partial
            \Delta\mu(z)}{\partial z}. \label{eq:chargecur}
\end{align}
The measured resistivity at some electric field strength $E$ is then given by
the ratio of the electric field and the averaged charge current:\begin{align}
    \rho(E)&=\frac{E}{\frac{1}{L_\mathrm{HM}}\int_{-L_\mathrm{HM}}^0\dif z\,
            j_\mathrm{c}(z)}. \label{eq:rho}
\end{align}
Finally, we define the USMR $\mathcal{U}$ as the fractional difference in
resistivity on inverting the electric field:\begin{align*}
    \mathcal{U}&\equiv\left|\frac{\rho(E)-\rho(-E)}{\rho(E)}\right|
        =\left|1+\frac{\int_{-L_\mathrm{HM}}^0\dif z\,j_\mathrm{c}(z; E)}{
            \int_{-L_\mathrm{HM}}^0\dif z\,j_\mathrm{c}(z; -E)}\right|.
\end{align*}

It should be noted that the even-ordered terms in the expansion of
the interface current are vital to the appearance of unidrectional SMR.
Suppose our system has equal magnon and electron temperature, such that the
interfacial spin Seebeck term $I_0$ vanishes (see
Section~\ref{subsec:thermal}), and we ignore the quadratic terms in
Eq.~(\ref{eq:intcur-exp}). Then because the only term in the spin current
equations~(\ref{eq:spincur}) that is independent of the accumulations is
$-\frac{\hbar\sigma\theta_\mathrm{SH}}{2e}E$ in Eq.~(\ref{eq:jse}), we
have that $\Delta\mu\propto\mu_\mathrm{m}\propto E$. Then by
Eqs.~(\ref{eq:chargecur}) and~(\ref{eq:rho}), $j_\mathrm{c}\propto E$ and
$\rho(E)\propto\frac{E}{E}$, such that $\mathcal{U}=0$. Conversely, with
quadratic terms in the interfacial spin current, $\rho(E)\sim\frac{E}{E+E^2}$,
and likewise if $I_0$ does not vanish, $\rho(E)\sim\frac{E}{1+E}$. Both cases
give nonvanishing USMR. Physically, one can say that the spin-dependent
electron and magnon populations couple together in a nonlinear fashion
(namely, through the Bose-Einstein distributions in Eq.~(\ref{eq:intcur})),
leading to a nonlinear dependence on the electric field.

\section{Results}
\label{sec:results}
\subsection{Equal-temperature, finite gap case}
\label{subsec:eqt}
Although our model can be solved analytically (up to evaluation of the
integrals $I_i$), the full expression of $\mathcal{U}$ is unwieldy and
therefore hardly insightful. To get an idea of the behavior of a real system,
we use a set of parameters---listed in
Table~\ref{tab:PtYIG-params}---corresponding to a Pt|YIG bilayer as a starting
point. (Unless otherwise specified, all parameters used henceforth are to be
taken from this table.)

\begin{figure}[h]
    \centering
    \includegraphics[width=\linewidth]{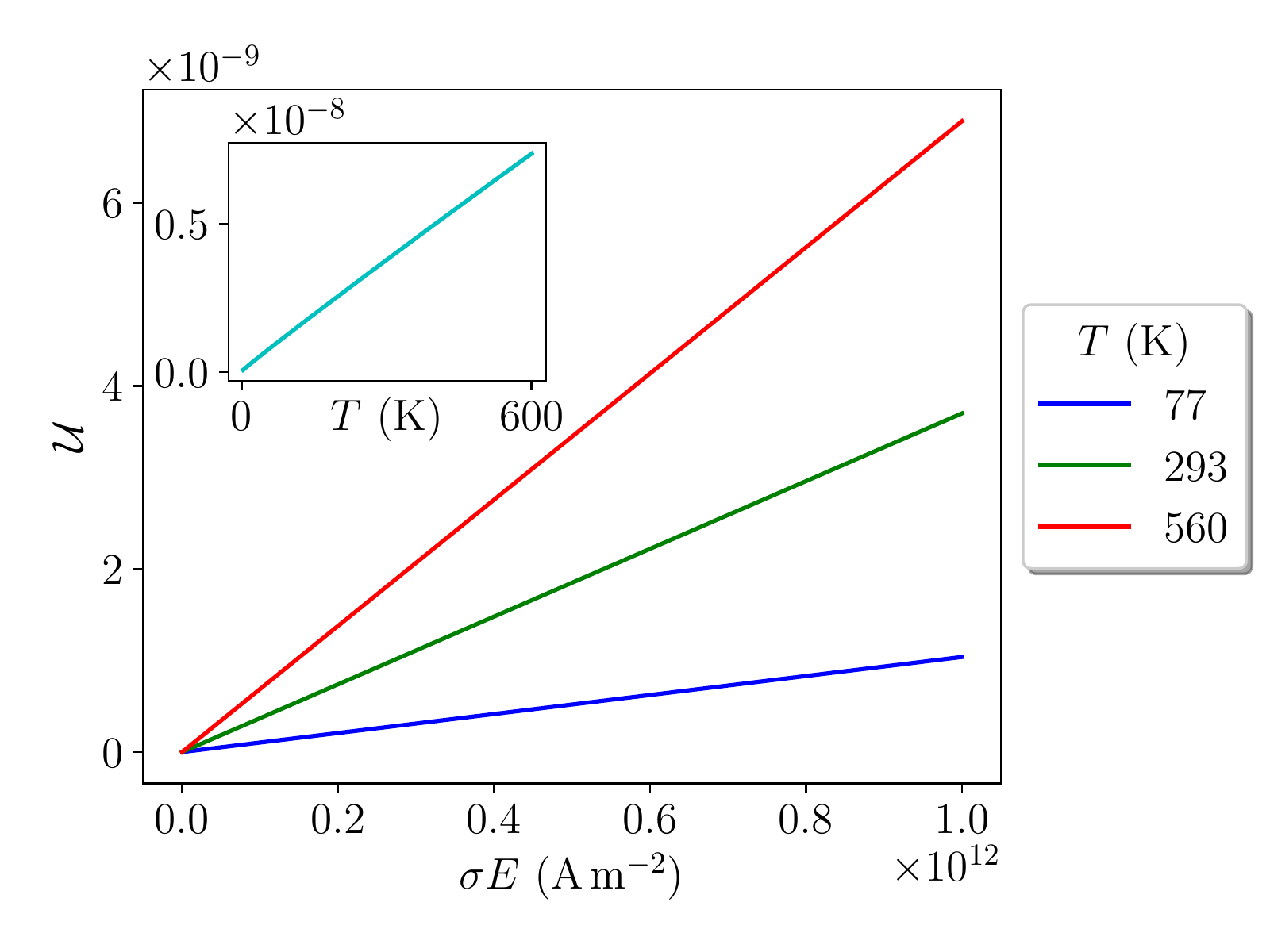}
    \caption{USMR $\mathcal{U}$ versus driving current $\sigma E$ for a Pt|YIG
    bilayer at liquid nitrogen temperature (\SI{77}{\kelvin}, blue), room
    temperature (\SI{293}{\kelvin}, green) and the YIG Curie temperature
    (\SI{560}{\kelvin}, red). Inset: USMR versus system temperature $T$ at
    fixed current $\sigma E=\SI{1e12}{\ampere\per\square\metre}$.}
    \label{fig:USMR-E}
\end{figure}

Fig.~\ref{fig:USMR-E} shows the magnonic USMR of a Pt|YIG bilayer versus
applied driving current ($\sigma E$) when $T_\mathrm{m}=T_\mathrm{e}=T$, at
the temperature of liquid nitrogen (\SI{77}{\kelvin}, blue), room temperature
(\SI{293}{\kelvin}, green) and the Curie temperature of YIG (\SI{560}{\kelvin}
\cite{1993PhR...229...81C}, red). FI and HM layer thicknesses used are
\SI{90}{\nano\metre} and \SI{3}{\nano\metre}, respectively, in line with
experimental measurements by \citet{2015ApPhL.107s2405A}.

In all cases the magnonic USMR is proportional to the applied electric
current---that is, the cubic term found by \citet{2018arXiv180605305O} is
absent---and at room temperature has a value on the order of $10^{-9}$ at
typical measurement currents \cite{2015NatPh..11..570A}. This is roughly four
orders of magnitude weaker than the USMR obtained---both experimentally and
theoretically---for FM|HM hybrids
\cite{2015NatPh..11..570A,2015ApPhL.107s2405A,2016PhRvB..94n0411Z,2017ApPhL.111w2405Y},
and is consistent with the experimental null results obtained for this system
by \citet{2015ApPhL.107s2405A}. Note, however, that the thickness of the FI
layer used by these authors is significantly lower than the magnon spin
diffusion length $l_\mathrm{m}=\SI{326}{\nano\metre}$, which results in a
suppressed USMR.

Furthermore, it can be seen in the inset of Fig.~\ref{fig:USMR-E} that the
magnonic USMR is, to good approximation, linear in the system temperature, in
agreement with observations by \citet{2016arXiv160308746K} and
\citet{2018arXiv180605305O}.

\begin{figure}[hb]
    \centering
    \includegraphics[width=.85\linewidth]{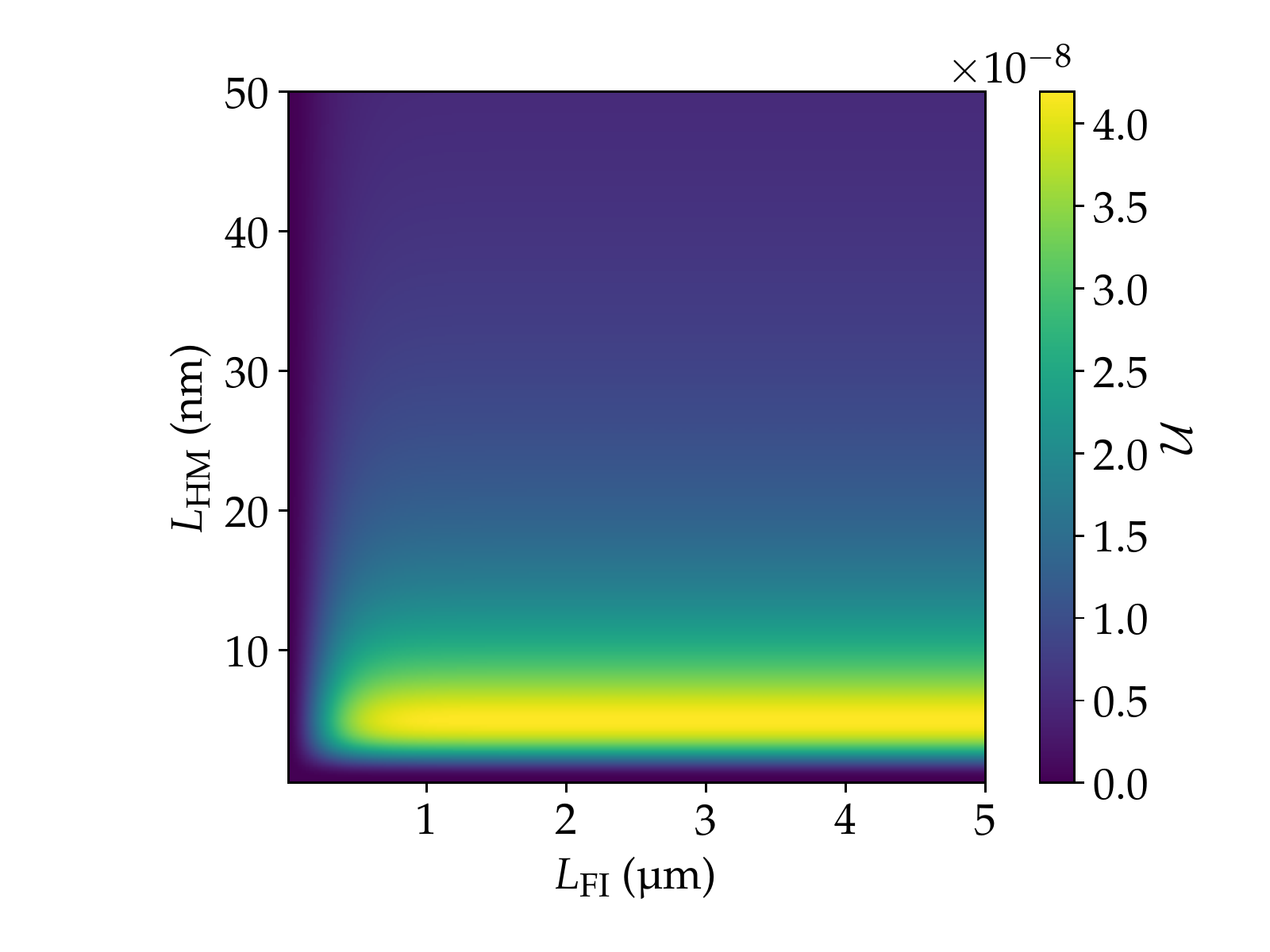}
    \caption{Pt|YIG USMR $\mathcal{U}$ at
    $T_\mathrm{m}=T_\mathrm{e}=\SI{293}{\kelvin}$ versus FI layer thickness
    $L_\mathrm{FI}$ and HM layer thickness $L_\mathrm{HM}$. A driving current
    $\sigma E=\SI{1e12}{\ampere\per\metre\squared}$ is used. A maximal USMR
    of \num{4.2e-8} is reached at $L_\mathrm{HM}=\SI{4.5}{\nano\metre}$,
    $L_\mathrm{FI}=\SI{5}{\micro\metre}$.}
    \label{fig:thickness}
\end{figure}

In Fig.~\ref{fig:thickness} we compute the USMR at
$\sigma E=\SI{1e12}{\ampere\per\metre\squared}$ as a function of both
$L_\mathrm{FI}$ and $L_\mathrm{HM}$. A maximum is reached around
$L_\mathrm{HM}\approx\SI{4.5}{\nano\metre}$, while in terms of
$L_\mathrm{FI}$, a plateau is approached within a few spin diffusion lengths.
By varying the layer thicknesses, a maximal USMR of \num{4.2e-8} can be
achieved, an improvement of one order of magnitude compared to the thicknesses
used by \citet{2015ApPhL.107s2405A}.

\subsection{Thermal effects}
\label{subsec:thermal}
We take into account a difference between the electron and magnon temperatures
$T_\mathrm{e}$ and $T_\mathrm{m}$ by assuming these parameters are equal to
the temperatures of the HM and FI layers, respectively, which we take to be
homogeneous. We assume that the HM undergoes ohmic heating and dissipates this
heat into the ferromagnet, which we take to be an infinite heat bath at
temperature $T_\mathrm{m}$. We only take into account the interfacial
(Kapitza) thermal resistance $R_\mathrm{th}$ between the HM and FI layers,
leading to a simple expression for the HM temperature
$T_\mathrm{e}$:\begin{align*}
    T_\mathrm{e}&=T_\mathrm{m}+R_\mathrm{th}\sigma E^2L_\mathrm{HM}.
\end{align*}

Using this model, we still find a linear dependence in the electric field,
$\mathcal{U}\simeq u_E(T_\mathrm{m}) \sigma E$, but the coefficient
$u_E(T_\mathrm{m})$ increases by three orders of magnitude compared to the
case where the electron and magnon temperatures are set to be equal. The
overwhelming majority of this increase can be attributed to an interfacial
spin Seebeck effect (SSE) \cite{2016PhRvB..94a4412C,2013PhRvB..88i4410S}: it is
caused by the accumulation-independent contribution $I_0$
(Eq.~(\ref{eq:I0})) in the interface current. When $I_0$ is artificially
set to 0, $u_E(T_\mathrm{m})$ changes less than 1\% from its equal-temperature
value.

Furthermore, the overall magnitude of the interfacial SSE in our system can be
attributed to the fact that we have a conductor|insulator interface: the
current runs through the HM only, resulting in inhomogeneous Joule heating of
the sample and a large temperature discontinuity across the interface.

\subsection{Spin Hall angle}
\label{subsec:theta}
\begin{figure}[h]
    \centering
    \includegraphics[width=.85\linewidth]{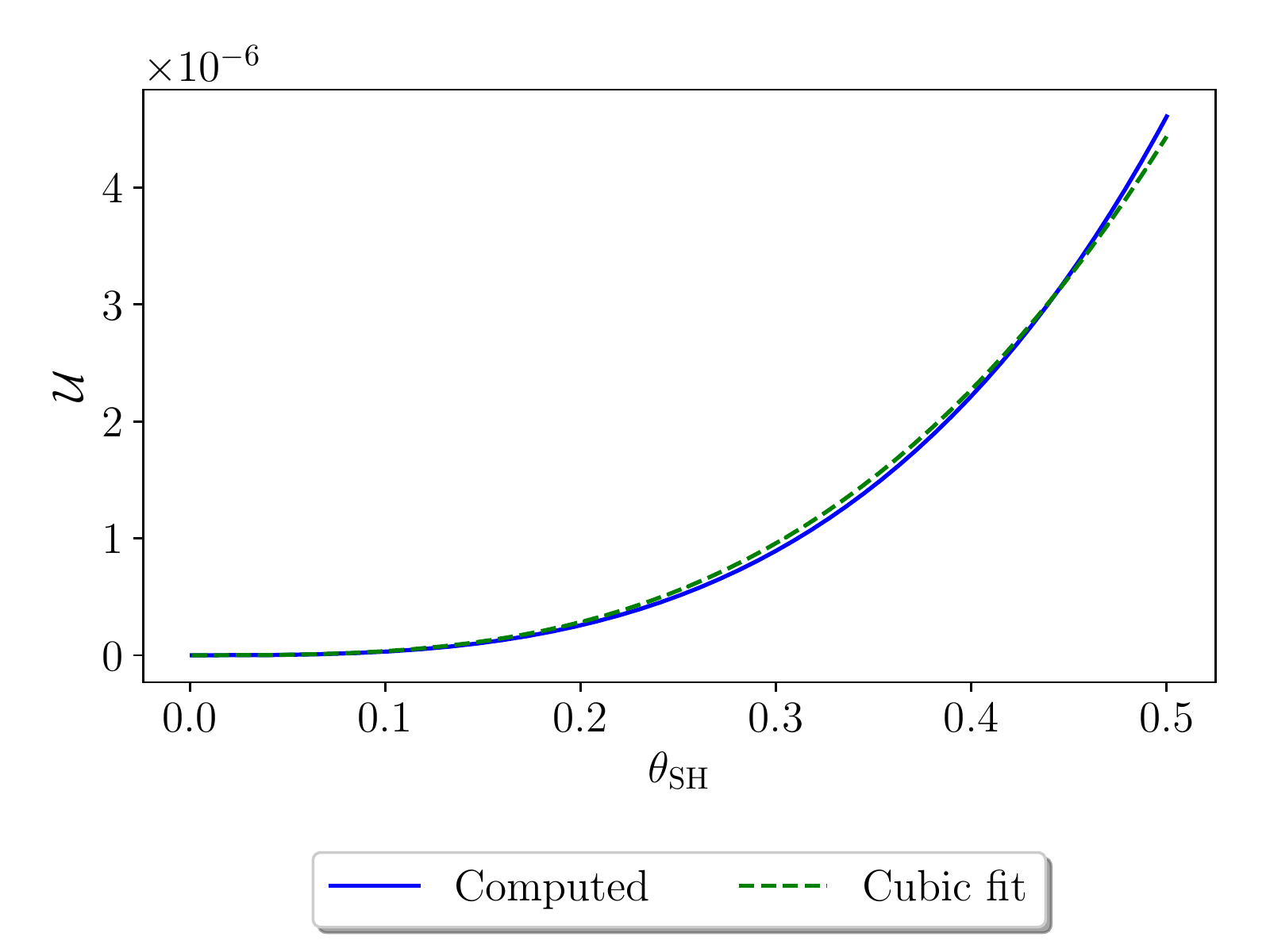}
    \caption{USMR $\mathcal{U}$ at
    $T_\mathrm{m}=T_\mathrm{e}=\SI{293}{\kelvin}$ versus spin Hall angle
    $\theta_\mathrm{SH}$. A driving current
    $\sigma E=\SI{1e12}{\ampere\per\metre\squared}$ and FI and HM layer
    thicknesses $L_\mathrm{FI}=\SI{5}{\micro\metre}$ and
    $L_\mathrm{HM}=\SI{4.5}{\nano\metre}$ are used. Blue curve: computed
    value. Dashed green curve: fit of the form
    $\mathcal{U}=u_\theta\theta_\mathrm{SH}^3$, with
    $u_\theta\simeq\num{3.1e-4}$.}
    \label{fig:theta}
\end{figure}

The electronic spin accumulation $\Delta\mu$ at the interface in the standard
spin Hall effect is linear in the electric field $E$ and spin Hall angle
$\theta_\mathrm{SH}$ \cite{2013PhRvB..87n4411C}.
From the linearity in $E$, we may conclude that the terms in
Eq.~(\ref{eq:intcur-exp}) that are linear in $\Delta\mu$ have a suppressed
contribution to the USMR. Thus, the contribution of the interface current is
of order $\theta_\mathrm{SH}^2$. Furthermore, $\Delta\mu$ enters the charge
current (Eq.~(\ref{eq:chargecur})) with a prefactor $\theta_\mathrm{SH}$,
leaving the magnonic USMR predominantly cubic in the spin Hall angle. Indeed,
in the special case $T_\mathrm{m}=T_\mathrm{e}$, expanding the full expression
for $\mathcal{U}$ (which spans several pages and is therefore not reproduced
within this work) in $\theta_\mathrm{SH}$ reveals that the first nonzero
coefficient is that of $\theta_\mathrm{SH}^3$. This suggests a small change in
$\theta_\mathrm{SH}$ potentially has a large effect on the USMR.

In Fig.~\ref{fig:theta} we plot the USMR for a Pt|YIG bilayer---once again
using $T_\mathrm{m}=T_\mathrm{e}=\SI{293}{\kelvin}$---consisting of
\SI{4.5}{\nano\metre} of Pt and \SI{5}{\micro\metre} of YIG, in which we
sweep the spin Hall angle. Included is a cubic fit
$\mathcal{U}=u_\theta\theta_\mathrm{SH}^3$, where we find
$u_\theta\simeq\num{3.1e-4}$. Here it can be seen that the magnonic USMR in
HM|FI bilayers can, as expected, potentially acquire magnitudes roughly
comparable to those in HM|FM systems, provided one can find or engineer a
metal with a spin Hall angle several times greater than that of Pt. This
suggests that very strong spin-orbit coupling (SOC) is liable to produce
significant magnon-mediated USMR in FI|HM heterostructures, although we
expect our model to break down in this regime.

\subsection{A note on the magnon spin diffusion length}
\label{subsec:lm}
Although we use the analytic expression for the magnon spin diffusion
length\cite{2012PhRvL.109i6603Z,2012PhRvB..86u4424Z,2016PhRvB..94a4412C},
\begin{align*}
    l_\mathrm{m}&=v_\mathrm{th}\sqrt{\frac{2}{3}\tau\tau_\mathrm{mr}}
\end{align*}
---where $v_\mathrm{th}$ is the magnon thermal velocity, $\tau$ is the
combined relaxation time, and $\tau_\mathrm{mr}$ is the magnonic relaxation
time (see Table~\ref{tab:PtYIG-params})---this is known to correspond poorly
to reality, being at least an order of magnitude too
low in the case of YIG \cite{2016PhRvB..94a4412C}. Artificially setting the
magnon spin-diffusion length to the experimental value of
\SI{10}{\micro\metre} (while otherwise continuing to use the parameters from
Table~\ref{tab:PtYIG-params}) results in a drop in USMR of some 4 orders of
magnitude.

\begin{figure}[ht]
    \centering
    \includegraphics[width=\linewidth]{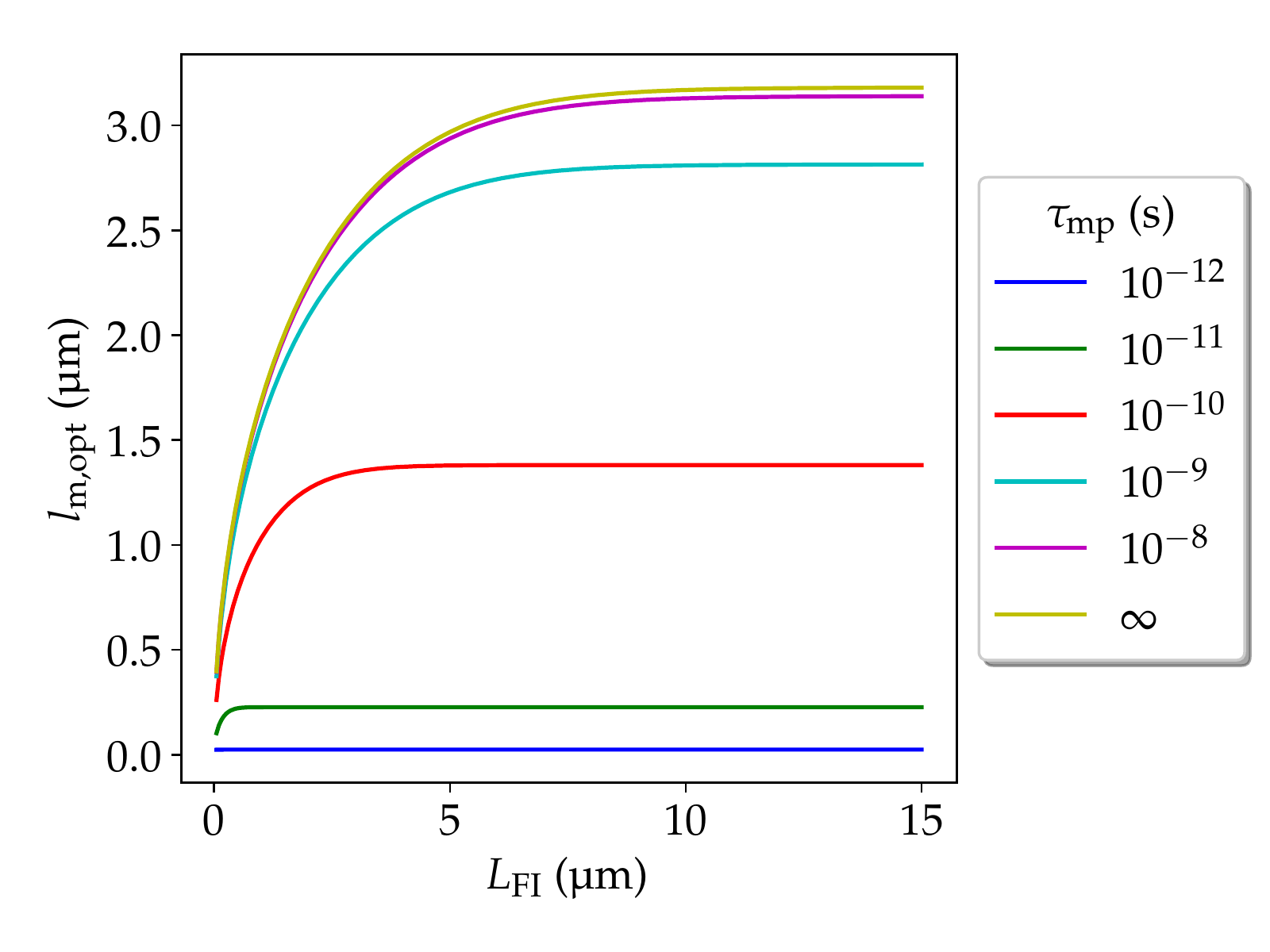}
    \caption{Value of the magnon spin diffusion length $l_\mathrm{m}$ that
    maximizes the USMR, as a function of FI layer thickness $L_\mathrm{FI}$, at
    various values of the magnon-phonon relaxation time $\tau_\mathrm{mp}$.}
    \label{fig:opt-lm}
\end{figure}

It follows directly that there exists some optimal value of $l_\mathrm{m}$
(which we shall label $l_\mathrm{m,opt}$) that maximizes the USMR, which we
plot as a function of the FI layer thickness $L_\mathrm{FI}$ in
Fig.~\ref{fig:opt-lm}, at $L_\mathrm{HM}=\SI{4.5}{\nano\metre}$ and $\sigma
E=\SI{1e12}{\ampere\per\metre\squared}$, and for various values of the
magnon-phonon relaxation time $\tau_\mathrm{mp}$, which is the shortest and
therefore most important timescale we take into account. For the physically
realistic value of $\tau_\mathrm{mp}=\SI{1}{\pico\second}$ (blue curve), the
optimal magnon spin diffusion length is just \SI{24}{\nano\metre}. Although
$l_\mathrm{m,opt}$ itself depends on $\tau_\mathrm{mp}$, the condition
$l_\mathrm{m}=l_\mathrm{m,opt}$ acts to cancel the dependence of the USMR on
the magnon-phonon relaxation time. Curiously, the USMR additionally loses its
dependence on $L_\mathrm{FI}$, reaching a fixed value of \num{4.14e-7} for our
parameters.

We further find that $l_\mathrm{m,opt}$ is independent of the spin Hall angle
and driving current, and shows a weak decrease with increasing temperature
provided the magnon-phonon scattering time is sufficiently short. A
significant increase in the optimal spin diffusion length is only found at low
temperatures and large $\tau_\mathrm{mp}$. Similarly, a weak dependence on the
Gilbert damping constant $\alpha$ is found, becoming more significant at large
$\tau_\mathrm{mp}$, with lower values of $\alpha$ corresponding to larger
$l_\mathrm{m,opt}$. When $\alpha$ is swept, again the USMR at
$l_\mathrm{m}=l_\mathrm{m,opt}$ acquires a universal value of \num{4.14e-7}
for our system parameters.

\subsection{Effect of the magnon gap}
\label{subsec:gap}
We have thus far utilised a fixed magnon gap with a value of
$\Delta/\mu_\mathrm{B}=\SI{1}{\tesla}$ for YIG. Although this is reasonable
for typical systems, it is possible to significantly reduce the gap size by
minimizing the anisotropy fields within the sample, e.g. using
a combination of external fields \cite{2005PhRvB..72a4454P}, optimized
sample shapes \cite{2016ApPhL.108j2403T,2015PhRvL.115w7201S} and temperature
\cite{1957PhRv..105..759D,1960JAP....31S.376R}. This leads us to consider the
effect a decreased or even vanishing gap may have on our results.

\begin{figure}
    \centering
    \includegraphics[width=\linewidth]{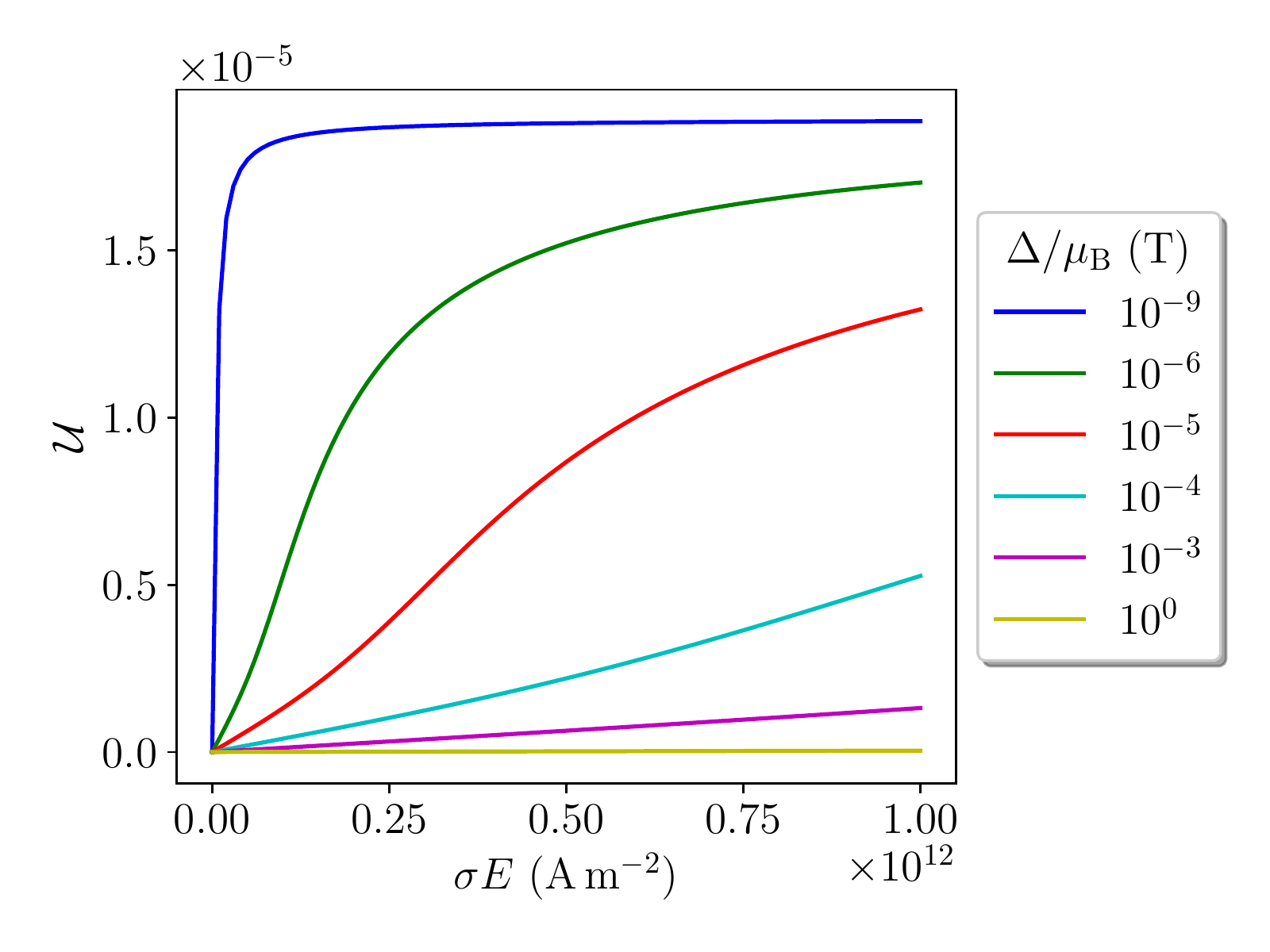}
    \caption{USMR $\mathcal{U}$ of a
    Pt(\SI{4.5}{\nano\metre})|YIG(\SI{5}{\micro\metre}) bilayer at room
    temperature versus applied current $\sigma E$ at various values of the
    magnon gap $\Delta$. For large gaps, linear behavior is recovered at
    realistic currents, while for smaller gap sizes, the USMR saturates as the
    current is increased.}
    \label{fig:E-delta}
\end{figure}

Fig.~\ref{fig:E-delta} shows the USMR $\mathcal{U}$ for a Pt|YIG system
(\SI{4.5}{\nano\metre} of Pt and \SI{5}{\micro\metre} of YIG) at room
temperature, plotted against the driving current $\sigma E$, now for different
values of the magnon gap $\Delta$. Here it can be seen that while
$\mathcal{U}$ is linear in $E$ for large gap sizes and realistic currents, it
shows limiting behavior at smaller gaps, becoming independent of the electric
current above some threshold (provided one neglects the effect of Joule
heating). At low current and intermediate magnon gap, the current dependence
is nonlinear at $\mathcal{O}(I^2)$ as opposed to the $\mathcal{O}(I^3)$
behavior found by \citet{2018arXiv180605305O}.

Note also that the saturation value of the USMR is two to three orders of
magnitude greater than the values found previously in our work, and of the
same magnitude as the electronic contribution found by
\citet{2016PhRvB..94n0411Z}.

The maximal value of the USMR that can be achieved may be found by considering
the full analytic expression for $\mathcal{U}$ in terms of the generic
coefficients $I_i$ representing the dimensionless integrals given by
Eqs.~(\ref{eq:intcur-coeffs}) in the Appendix. In the gapless limit
$\Delta\to0$ and at equal magnon and electron temperature
($T_\mathrm{m}=T_\mathrm{e}$), the second-order coefficients $I_\mathrm{mm}$
and $I_\mathrm{me}$ diverge, while their sum takes the constant value
$\lambda\equiv I_\mathrm{mm}+I_\mathrm{me}\simeq0.323551$ at room temperature.
$I_\mathrm{ee}$ does not diverge, and obtains the value $-\lambda$.

Now working in the thick-ferromagnet limit ($L_\mathrm{FI}\to\infty$), we
substitute $I_\mathrm{me}\to-I_\mathrm{mm}+\lambda$ and take the limits
$E\to\infty$ and $I_\mathrm{mm}\to-\infty$. By application of l'Hôpital's rule
in the latter, all coefficients $I_i$ drop out of the expression for
$\mathcal{U}$. This leaves only the asymptotic value, which, after expanding
in $\theta_\mathrm{SH}$, reads\begin{align}
    \mathcal{U}_\text{max}\!&=\!\frac{4e^2l_\mathrm{s}^2
            \theta_\mathrm{SH}^2\sigma_\mathrm{m}\tanh^2\left(\!
            \frac{L_\mathrm{HM}}{2l_\mathrm{s}}\!\right)}{\hbar^2l_\mathrm{m}
            L_\mathrm{HM}\sigma\!+\!4l_\mathrm{s}e^2L_\mathrm{HM}
            \sigma_\mathrm{m}\coth\left(\!\frac{L_\mathrm{HM}}{l_\mathrm{s}}
            \!\right)}\!+\!\mathcal{O}(\theta_\mathrm{SH}^4). \label{eq:lim}
\end{align}
Whereas the linear-in-$E$ regime of the magnonic USMR grows
as $\theta_\mathrm{SH}^3$, we thus find that the leading-order behavior of the
\emph{asymptotic} value is only $\theta_\mathrm{SH}^2$, and the third-order
term vanishes completely. Physically, this can be explained by the fact that
the asymptotic magnonic USMR is purely a bulk effect: all details about the
interface vanish, while parameters originating from the bulk spin- and charge
currents remain. The appearance of $l_\mathrm{m}$ in the denominator and its
absence in the numerator of Eq.~(\ref{eq:lim}) once again highlights that a
large magnon spin diffusion length acts to \emph{suppress} the USMR.

\begin{figure}
    \centering
    \includegraphics[width=\linewidth]{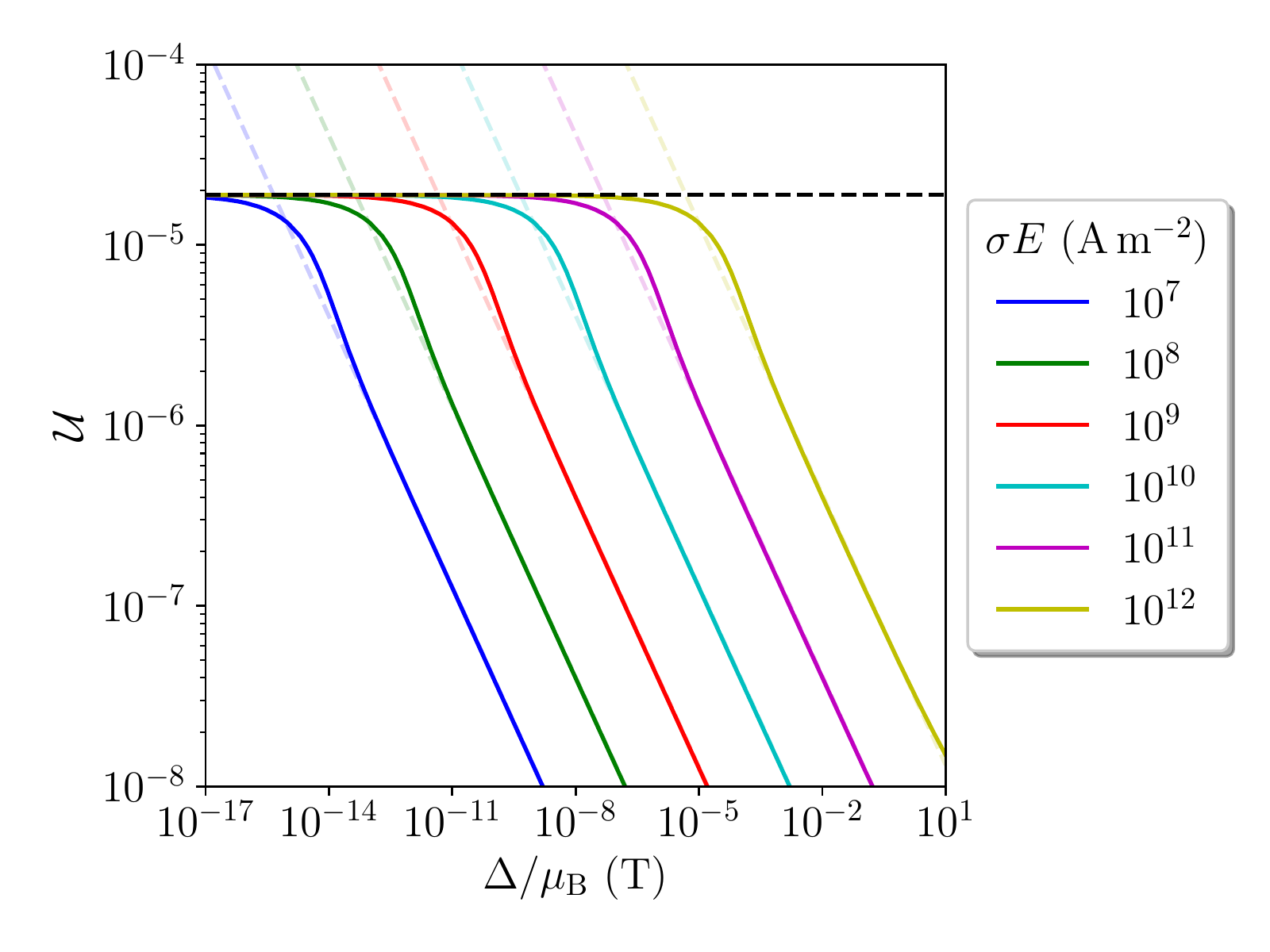}
    \caption{USMR $\mathcal{U}$ of a
    Pt(\SI{4.5}{\nano\metre})|YIG(\SI{5}{\micro\metre}) bilayer at room
    temperature as function of the magnon gap size $\Delta$, for various
    values of the base charge current $\sigma E$.  Note the log-log scaling.
    Solid colored lines: computed USMR. Dashed colored lines: continuations of
    the high-gap tails of the corresponding curves according to the
    one-parameter fit $\mathcal{U}=u_0/\sqrt{\Delta}$. Dashed black line:
    asymptotic value of the USMR as given by Eq.~(\ref{eq:lim}).}
    \label{fig:delta}
\end{figure}

Fig.~\ref{fig:delta} is a log-log plot of the USMR versus gap size $\Delta$ at
various values of the driving current $\sigma E$. Here the value
$\mathcal{U}_\text{max}$ is shown as a dashed black line, indicating that this
is indeed the value to which $\mathcal{U}$ converges in the gapless limit or
at high current. Moreover, it shows that for given $\sigma E$, one can find a
turning point at which the USMR switches relatively abruptly from being nearly
constant to decreasing as $1/\sqrt{\Delta}$.

A (backwards) continuation of the decreasing tails is included in
Fig.~\ref{fig:delta} as dashed lines following the one-parameter fit
$\mathcal{U}=u_0/\sqrt{\Delta}$, and we define the threshold gap
$\Delta_\mathrm{th}$ as the value of $\Delta$ where this continuation
intersects $\mathcal{U}_\text{max}$. We then find that $\Delta_\mathrm{th}$
scales as $E^2$, or conversely, that the driving current required to saturate
the USMR scales as the square root of the magnon gap.

We note that although the small-gap regime is mathematically valid (even in
the limit $\Delta\to0$, as $\Delta$ may be brought arbitrarily close to 0 in a
continuous manner), it does not necessarily correspond to a physical
situation: when the anisotropy vanishes, the magnetization of the FI layer may
be reoriented freely, which will break our initial assumptions. Nevertheless,
in taking the gapless limit, we are able to predict an upper limit on the
magnonic USMR.

\section{Conclusions}
\label{sec:conclusion}
Using a simple drift-diffusion model, we have shown that magnonic spin
accumulation near the interface between a ferromagnetic insulator and a heavy
metal leads to a small but nonvanishing contribution to the unidirectional
spin Hall magnetoresistance of FI|HM heterostructures. Central to our model is
an interfacial spin current originating from a spin-flip scattering process
whereby electrons in the heavy metal create or annihilate magnons in the
ferromagnet. This current is markedly nonlinear in the electronic and magnonic
spin accumulations at the interface, and it is exactly this nonlinearity which
gives rise to the magnonic USMR.

For Pt|YIG bilayers, we predict that the magnonic USMR $\mathcal{U}$ is at
most on the order of $10^{-8}$, roughly three orders of magnitude weaker than
the measured USMR in FM|HM hybrids (where electronic spin accumulation is
thought to form the largest contribution). This is fully consistent with
experiments that fail to detect USMR in Pt|YIG systems, as the tiny signal is
drowned out by the interfacial spin Seebeck effect, which has a similar
experimental signature and is enhanced compared to the FM|HM case due to
inhomogeneous Joule heating.

We have shown that the magnon-mediated USMR is approximately cubic in the
spin Hall angle of the metal, suggesting that metals with extremely large spin
Hall angles may provide a significantly larger USMR than Pt. It is
therefore plausible that a large magnonic USMR can exist in systems with very
strong spin-orbit coupling, even though our model would break down in this
regime.

The magnonic USMR depends strongly on the magnon spin diffusion length
$l_\mathrm{m}$ in the ferromagnet. Motivated by a large discrepancy between
experimental values and theoretical predictions of $l_\mathrm{m}$, we have
shown that a significant increase in USMR can be realized if a method is found
to engineer this parameter to specific, optimal values that, for realistic
values of the magnon-phonon relaxation time $\tau_\mathrm{mp}$ (on the order
of \SI{1}{\pico\second} for YIG), are significantly shorter than those
measured experimentally or computed theoretically. We further find that when
the magnon spin diffusion length has its optimal value, the USMR becomes
independent of the ferromagnet's thickness and Gilbert damping constant.

Although in physically reasonable regimes, the magnonic USMR is to very good
approximation linear in the applied driving current $\sigma E$, it saturates
to a fixed value given extremely large currents or a strongly reduced magnon
gap $\Delta$. The transition from linear to constant behavior in the driving
current is heralded by a turning point which is proportional to the square
root of the magnon gap. The asymptotic behavior of the USMR beyond the turning
point is governed by the bulk spin- and charge currents, and is completely
independent of the details of the interface.

While a vast reduction in $\Delta$ is required to bring the saturation current
of a Pt|YIG bilayer within experimentally reasonable regimes, the magnonic
USMR scales as $1/\sqrt{\Delta}$ at currents below the turning point,
suggesting that highly isotropic FI|HM samples are most likely to produce a
measurable magnonic USMR. The increase in magnonic USMR at low gaps (and large
currents) is in good qualitative agreement with the recent experimental work
of \citet{2018arXiv180605305O}, as is the linear dependence on system
temperature.

A notable disagreement with the experimental data of
\citet{2018arXiv180605305O} is found in the scaling of the current dependence,
which in our results lacks an $\mathcal{O}(I^3)$ term at large magnon gaps and
contains an $\mathcal{O}(I^2)$ term at intermediate gaps. It is still unclear
whether this discrepancy can be explained by system differences, such as the
finite electrical resistance of Co or the presence of Joule heating.

Finally, we note that while our results apply to ferromagnetic insulators, it
is reasonable to assume a magnonic contribution also exists in HM|FM
heterostructures, although the possibility of coupled transport of magnons and
electrons makes such systems more difficult to model.  Additionally, various
extensions of our model may be considered, such as the incorporation of
spin-momentum locking \cite{2016PhRvL.117l7202Y}, ellipticity of magnons, heat
transport and nonuniform temperature profiles \cite{2016PhRvB..94a4412C},
directional dependence of the magnetization, etc.

\section{Acknowledgements}
R.A.D. is member of the D-ITP consortium, a program of the Dutch Organisation
for Scientific Research (NWO) that is funded by the Dutch Ministry of
Education, Culture and Science (OCW). This work is funded by the European
Research Council (ERC).

\bibliographystyle{apsrev4-1}
\bibliography{refs}

\FloatBarrier
\begin{table*}
    \centering
    \begin{tabular}{|lllrr|}
        \hline
        Description
        & Symbol
        & Expression
        & Value at $T=\SI{293}{\kelvin}$
        & Ref.
        \\ \hline
        YIG spin-wave stiffness constant
        & $J_\mathrm{s}$
        &
        & \SI{8.458e-40}{\joule\metre\squared}
        & \cite{2016PhRvB..94a4412C} \\
        YIG spin quantum number per unit cell
        & $S$
        &
        & 10
        & \cite{2016PhRvB..94a4412C} \\
        YIG lattice constant
        & $a$
        &
        & \SI{1.2376}{\nano\metre}
        & \cite{2016PhRvB..94a4412C} \\
        YIG Gilbert damping constant
        &$\alpha$
        &
        & \num{1e-4}
        & \cite{2016PhRvB..94a4412C} \\
        YIG spin number density
        & $s$
        & $S a^{-3}$
        & \SI{5.2754e27}{\per\metre\cubed}
        & \cite{2016PhRvB..94a4412C} \\
        YIG magnon gap
        & $\Delta$
        &
        & \SI{9.3e-24}{\joule}
        & \cite{1993PhR...229...81C} \\
        YIG magnon-phonon scattering time
        & $\tau_\mathrm{mp}$
        &
        & \SI{1}{\pico\second}
        & \cite{2016PhRvB..94a4412C} \\
        YIG magnon relaxation time
        & $\tau_\mathrm{mr}$
        & $\frac{\hbar}{2\alpha\kBT_\mathrm{m}}$
        & \SI{130}{\pico\second}
        & \cite{2016PhRvB..94a4412C} \\
        Combined magnon relaxation time
        & $\tau$
        & $\left(\frac{1}{\tau_\mathrm{mr}}+\frac{1}{\tau_\mathrm{mp}}
            \right)^{-1}$
        & \SI{1}{\pico\second}
        & \cite{2016PhRvB..94a4412C} \\
        Magnon thermal de Broglie wavelength
        & $\Lambda$
        & $\sqrt{\frac{4\uppi J_\mathrm{s}}{\kBT_\mathrm{m}}}$
        & \SI{1.62}{\nano\metre}
        & \cite{2016PhRvB..94a4412C} \\
        Magnon thermal velocity
        & $v_\mathrm{th}$
        & $\frac{2\sqrt{J_\mathrm{s}\kBT}}{\hbar}$
        & \SI{35.1}{\kilo\metre\per\second}
        & \cite{2016PhRvB..94a4412C} \\
        Magnon spin diffusion length
        & $l_\mathrm{m}$
        & $v_\mathrm{th}\sqrt{\frac{2}{3}\tau\tau_\mathrm{mr}}$
        & \SI{326}{\nano\metre}
        & \cite{2016PhRvB..94a4412C} \\
        Magnon spin conductivity
        & $\sigma_\mathrm{m}$
        & $\zeta\left(\frac{3}{2}\right)^2\frac{J_\mathrm{s}}{\Lambda^3}\tau$
        & \SI{1.35e-24}{\joule\second\per\metre}
        & \cite{2016PhRvB..94a4412C} \\
        Real part of spin-mixing conductance
        & $g_\mathrm{r}^{\uparrow\downarrow}$
        &
        & \SI{5e18}{\per\metre\squared}
        & \cite{2012PhRvL.108x6601B} \\
        Pt electrical conductivity
        & $\sigma$
        &
        & \SI{1e7}{\siemens\per\metre}
        & \cite{asmhandbook}\footnote{The conductivity of Pt is approximately
        inverse-linear in temperature over the regime we are considering.
        However, as we are not interested in detailed thermodynamic behavior,
        we use the fixed value $\sigma=\SI{1e7}{\siemens\per\metre}$
        throughout this work.} \\
        Pt spin Hall angle
        & $\theta_\mathrm{SH}$
        &
        & 0.11
        & \cite{2016PhRvB..94a4412C} \\
        Pt electron diffusion length
        & $l_\mathrm{s}$
        &
        & \SI{1.5}{\nano\metre}
        & \cite{2016PhRvB..94a4412C} \\
        Pt|YIG Kapitza resistance
        & $R_\mathrm{th}$
        &
        & \SI{3.58e-9}{\metre\squared\kelvin\per\watt}
        & \cite{2013PhRvB..88i4410S} \\
        \hline
    \end{tabular}
    \caption{System parameters for a Pt|YIG bilayer film.}
    \label{tab:PtYIG-params}
\end{table*}

\begin{widetext}
\appendix*
\section{Interfacial spin current integrals}
\label{app:integrals}
The following dimensionless integrals appear in the second-order expansion of
the interfacial spin current to the spin accumulations,
Eq.~(\ref{eq:intcur-exp}):
\begin{subequations}\label{eq:intcur-coeffs}\begin{align}
    I_\mathrm{0}&=\int_{\frac{\Delta}{\kBT_\mathrm{m}}}^\infty\dif x\,\sqrt{x
            -\frac{\Delta}{\kBT_\mathrm{m}}}x\left(\bose{x}
            -\bose{\frac{T_\mathrm{m}}{T_\mathrm{e}} x}\right), \label{eq:I0}
            \\
    I_\mathrm{e}&=\int_{\frac{\Delta}{\kBT_\mathrm{m}}}^\infty\dif x\,\sqrt{x
            -\frac{\Delta}{\kBT_\mathrm{m}}}\left(\bose{\frac{T_\mathrm{m}}{
            T_\mathrm{e}} x}-\bose{x}-\frac{T_\mathrm{m}}{T_\mathrm{e}}
            x\upe^{\frac{T_\mathrm{m}}{T_\mathrm{e}} x}\bosepow{2}{
            \frac{T_\mathrm{m}}{T_\mathrm{e}} x}\right), \\
    I_\mathrm{m}&=\int_{\frac{\Delta}{\kBT_\mathrm{m}}}^\infty\dif x\,\sqrt{x
            -\frac{\Delta}{\kBT_\mathrm{m}}}x\upe^x\bosepow{2}{x}, \\
    I_\mathrm{ee}&=\int_{\frac{\Delta}{\kBT_\mathrm{m}}}^\infty\dif x\,
            \sqrt{x-\frac{\Delta}{\kBT_\mathrm{m}}}\left(
            \upe^{\frac{T_\mathrm{m}}{T_\mathrm{e}} x}\bosepow{3}{
            \frac{T_\mathrm{m}}{T_\mathrm{e}} x}
            \left[\upe^{\frac{T_\mathrm{m}}{T_\mathrm{e}} x}-1
            -\frac{T_\mathrm{m}x}{2T_\mathrm{e}}\left(
            \upe^{\frac{T_\mathrm{m}}{T_\mathrm{e}} x}+1\right)\right]\right),
            \\
    I_\mathrm{mm}&=\int_{\frac{\Delta}{\kBT_\mathrm{m}}}^\infty\dif x\,
            \sqrt{x-\frac{\Delta}{\kBT_\mathrm{m}}}\frac{x}{2}\upe^{x}
            \left[\upe^x+1\right]\bosepow{3}{x}, \\
    I_\mathrm{me}&=-\int_{\frac{\Delta}{\kBT_\mathrm{m}}}^\infty\dif x\,
            \sqrt{x-\frac{\Delta}{\kBT_\mathrm{m}}}\upe^x\bosepow{2}{x}.
\end{align}\end{subequations}
\end{widetext}
\end{document}